\documentclass{nature_preprint}
\usepackage[dvips]{graphics,color}
\title{A map of the day-night contrast of the extrasolar planet HD 189733b}

\author{Heather A. Knutson$^{1}$, David Charbonneau$^{1}$, Lori E. Allen$^{1}$, Jonathan J. Fortney$^{2,3}$, Eric Agol$^{4}$, Nicolas B. Cowan$^{4}$, Adam P. Showman$^{5}$, Curtis S. Cooper$^{5}$, \& S. Thomas Megeath$^{6}$}

\begin{document}

\maketitle

\begin{affiliations}

\item Harvard-Smithsonian Center for Astrophysics, 60 Garden Street, Cambridge, MA 02138, USA

\item Space Science and Astrobiology Division, NASA Ames Research Center, MS 245-3, Moffett Field, CA 94035, USA

\item SETI Institute, 515 N. Whisman Road, Mountain view, CA 94043, USA

\item Department of Astronomy, Box 351580, University of Washington, Seattle, WA 98195, USA

\item Lunar and Planetary Laboratory and Department of Planetary Sciences, University of Arizona, Tucson, AZ 85721, USA

\item Department of Physics and Astronomy, University of Toledo, 2801 West Bancroft St., Toledo, OH 43606, USA

\end{affiliations}

\begin{abstract}

``Hot Jupiter'' extrasolar planets are expected to be tidally locked because they are close ($<$0.05 astronomical units, where 1 AU is the average Sun-Earth distance) to their parent stars, resulting in permanent daysides and nightsides.  By observing systems where the planet and star periodically eclipse each other, several groups have been able to estimate the temperatures of the daysides of these planets\cite{1,2,3}.  A key question is whether the atmosphere is able to transport the energy incident upon the dayside to the nightside, which will determine the temperature at different points on the planet's surface.  Here we report observations of HD 189733, the closest of these eclipsing planetary systems\cite{4,5,6}, over half an orbital period, from which we can construct a 'map' of the distribution of temperatures.  We detected the increase in brightness as the dayside of the planet rotated into view.  We estimate a minimum brightness temperature of 973$\pm$33~K and a maximum brightness temperature of 1212$\pm$11~K at a wavelength of 8~$\mu$m, indicating that energy from the irradiated dayside is efficiently redistributed throughout the atmosphere, in contrast to a recent claim for another hot Jupiter\cite{7}.  Our data indicate that the peak hemisphere-integrated brightness occurs 16$\pm$6 degrees before opposition, corresponding to a hot spot shifted east of the substellar point.  The secondary eclipse (when the planet moves behind the star) occurs 120$\pm$24~s later than predicted, which may indicate a slightly eccentric orbit.  

\end{abstract}

We monitored HD 189733 continuously over a 33.1 hour period using the 8~$\mu$m channel of the InfraRed Array Camera (IRAC)\cite{8} on the Spitzer Space Telescope\cite{9}, observing in subarray mode with a cadence of 0.4~s.  Our observations spanned slightly more than half of the planet's orbit, beginning 2.6 hours before the start of the transit (when the planet moves in front of the star) and ending 1.9 hours after the end of the secondary eclipse.  This gave us a total of 278,528 $32 \times 32$ pixel images.  We found that there was a gradual detector-induced rise of up to 10\% in the signal measured in individual pixels over time.  This rise is illumination-dependent; pixels with high levels of illumination (greater than 250 MJy sr$^{-1}$) converge to a constant value within the first two hours of observations and lower-flux pixels increase linearly over time.  We characterize this effect by producing a time series of the signal in a series of annuli of increasing radius centered on the star (masking out a 5-pixel-wide box centered on HD 189733's smaller, fainter M dwarf companion\cite{10}).  This set of curves describes the behavior of the ramp for different illumination levels.  

To correct our images, we estimate the median illumination for each pixel in the array, and interpolate over our base set of curves (scaling as the natural log of the illumination) to calculate a curve describing the behavior of that pixel.  We correct for this instrumental effect by dividing the flux in each pixel in a given image by the value of the interpolated curve.  Pixels with illumination levels higher than 210 MJy sr$^{-1}$ are not corrected, as these pixels converge to a constant value before the transit.  We subsequently measure the flux from the M dwarf companion and find that it is constant at a level of $<0.05$\%, indicating that the detector effect has been removed.  We then use aperture photometry with a radius of 3.5 pixels to create our time series (see Fig. 1 for additional details).  We chose the smallest aperture possible while still avoiding flux losses from the shifting position of the star on the array; only 33\% of the total flux in our aperture comes from corrected pixels.  We test the effect of our correction on the size of the observed signal by inserting an artificial phase variation signal into the images before applying our corrections; we find that the amplitude of this signal is reduced by only 13\% of its total size.  

In our final time series (Fig. 1), we see a distinct rise in flux beginning shortly after the end of the transit and continuing until a time just prior to the beginning of the secondary eclipse.  (The transit and secondary eclipse occur at orbital phases $\sim$ 0 and $\sim$ 0.5, respectively.)  We estimate its amplitude by fitting a small region of the phase curve around the peak with a quadratic function, and taking the maximum value of that function as the peak of the phase curve.  After similarly fitting the region around the minimum, we estimate the total amplitude of this rise to be $0.12\pm0.02$\%, with uncertainties that are dominated by our correction for the detector effect (we propagate this systematic effect in all of our stated uncertainties below).  By comparing this variation to the depth of the secondary eclipse, we find that the minimum hemisphere-integrated flux is $64\pm7$\% of the maximum flux.  The peak in flux occurs $2.3\pm0.8$ hours before the center of the secondary eclipse, corresponding to a position $16\pm6$ degrees before opposition.  A possible confusing effect results from the fact that HD~189733 is an active star, with spots that cause the flux to vary by as much as $\pm1.5\%$ in visible light over its 13.4 day rotation period\cite{6}.  We estimate the size of this effect at 8~$\mu$m by treating both the star and the spots as blackbodies with temperatures of 5050~K and 4050~K respectively and scaling the variations observed at visible wavelengths to the appropriate 8~$\mu$m amplitude.  Projecting these variations forward in time, we find that there could be a linear increase in flux of 0.1\% over the period of our observations as the spots rotate into view.  Importantly, we note that accounting for these spots would serve only to reduce the amplitude of the observed phase variation.  As the shape of the observed variation is consistent with a genuine variation in the flux from the planet, we treat it as such in the discussions below.

The high quality of our data allows us to derive more precise estimates of the parameters for the planetary system than are currently available\cite{4,5,6} (see Fig. 2 and its legend for details).  We calculate a brightness temperature of $1205.1\pm9.3$~K for the dayside of the planet from the observed depth of the secondary eclipse.  We estimate that the planet has a minimum hemisphere-averaged brightness temperature of $973\pm33$~K occurring $6.7\pm0.4$ hours after the transit, and a maximum hemisphere-averaged brightness temperature of $1212\pm11$~K occurring $2.3\pm0.8$ hours before the onset of the secondary eclipse.

We find the center of the transit occurs at $t_{\rm I} = 2454037.611956 \pm 0.000067 $ HJD (6~s error), while the center of the secondary eclipse occurs at time $t_{\rm II} = 2454038.72294 \pm 0.00027$ HJD (24~s error), where the errors have been estimated using both a bootstrap method and a $10^5$ step Markov chain.  These are the most precise timing measurements of a transit and secondary eclipse to date.  The transit occurs at the predicted time\cite{6}, but the secondary eclipse occurs $150\pm24$~s later than its predicted time of half an orbital period after the transit.  Because we observe both eclipses and the period is well-constrained, we are able to predict the time of secondary eclipse with no significant uncertainty.  Part of the delay of the secondary eclipse is due to the light travel time across the system\cite{13} of 30~s.  The remaining delay is possibly due to (a) non-uniformity in the planet emission\cite{14,15}; (b) third bodies in the system; or (c) an eccentric orbit.  

To estimate the magnitude of the first effect, we fit the observed phase variation with a simple model of the planet consisting of twelve longitudinal slices of constant brightness.  The resulting model light curve is shown in Fig. 1b, and the best-fit longitudinal flux values are shown in Fig. 3b (see figure legend for more details).  Figure 3b is effectively a coarse 8~$\mu$m map of the planet with a resolution of 30 degrees in longitude and no resolution in latitude.  Figure 3a shows this brightness distribution projected onto the surface of the planet with an additional sinusoidal dependence on latitude included.  Because we observe the planet over only half an orbital period, the error bars are largest for longitudes near 90 degrees west of the substellar point.  Although this brightness distribution is a good fit for the later part of the phase curve (Fig. 1), a deviation is apparent near the transit; this fit could be improved by using a finer longitude resolution.  We find that the brightest slice on the planet is 30 degrees east of the sub-stellar point. The faintest slice of the planet also (surprisingly) appears in the eastern hemisphere, 30 degrees west of the antistellar point.  The brightest slice of the planet is roughly twice as bright as the faintest slice, corresponding to a temperature difference of $\sim$350~K.  This non-uniform brightness distribution changes the shape of the ingress and egress\cite{14,15}.  Treating the planet as a uniformly bright disk in our fit creates an artificial delay of at most 20~s in the time of secondary eclipse.  Thus, the planet's non-uniform emission cannot account for the 120~s delay of the secondary eclipse.

This offset is unlikely to be the result of perturbations to the planet's mean motion from a third body in the system; such perturbations would shift the time of the transit as well, and we see no evidence for such a shift.  This leaves the third option as the most likely explanation.  If the time delay is attributed to eccentricity $e$, then $e \cos{\varpi} = 0.0010 \pm 0.0002$ where $\varpi$ is the longitude of pericentre, indicating that the eccentricity is extremely small, but non-zero.  This is surprising, as the time scale for orbital circularization is significantly shorter than the ages of these systems\cite{16,17}.  This eccentricity is too small to have been detected by radial velocity measurements\cite{4,6}.  The observed delay is moderately inconsistent with the timing of the 16~$\mu$m eclipse\cite{3}, which occurs $29\pm65$~s later than predicted\cite{6}.

Atmosphere models allow us some insight into the factors that control the day-night temperature contrast.  The response of a planet to stellar irradiation depends on a comparison between the radiative time scale (over which starlight absorption and infrared emission alter the temperature) and the advection time scale (over which air parcels travel between day and night sides)\cite{18,19,20}.  If the radiative time is much shorter than the advection time, the hot day side reradiates the absorbed stellar flux and the night side remains cold.  If the radiative time greatly exceeds the advection time, however, then efficient thermal homogenization occurs.  Radiative transfer models of highly irradiated giant planets\cite{19,20,21,22,23} predict that the bulk of absorption of stellar flux and emission of thermal flux occurs at pressures from tens of millibars to several bars, where the predicted radiative time scales\cite{20} range from $10^4$--$10^5$ s.  Advection times are less well constrained, but estimates of wind speeds\cite{18,24,25,26,27,28} (hundreds to thousands of m~s$^{-1}$) suggest advection times of $\sim10^5$ s.  Thus, current models suggest that the radiative time scale is comparable to the advective time scale, and temperature differences could reach 1000~K.  In contrast, the small flux variation that we observe implies that the time scale for altering the temperature by radiation modestly exceeds the time scale for homogenizing the temperature between the day and night sides. 

It is possible that the observed planetary flux emerges from deeper in the atmosphere than expected, where the radiative time scales are longer.  In the 8~$\mu$m band, models suggest that H$_2$O dominates the opacity, with additional contributions from CH$_4$ and collision-induced absorption of H$_2$.  Silicate cloud opacity is not expected at these temperatures\cite{29}.  If the radiative time constants are as small as expected\cite{20} then supersonic wind speeds exceeding $\sim$10~km~s$^{-1}$ ($\sim 4$ times the sound speed) would be necessary to transport energy to the night side.  The times of minimum and maximum flux also provide information on the planet's meteorology.  Our observation that the minimum and maximum do not occur at phases of 0 and 0.5, respectively, indicates advection of the temperature pattern by atmospheric winds\cite{18,24,25,26,27,28,30}.  The existence of a flux minimum and maximum on a single hemisphere suggests a complex pattern not yet captured in current circulation models.

In contrast to the 8~$\mu$m phase variation for HD~189733b presented here, the 24~$\mu$m variation reported\cite{7} for the non-transiting planet $\upsilon$~And~b was quite large.  The reasons for the differing results are not immediately clear, although the sparse data sampling and unknown radius for $\upsilon$~And~b  mean that the uncertainty in the inferred day-night contrast is much larger.  A higher opacity at 24 $\mu$m and a lower surface gravity for $\upsilon$~And~b could lead to a photospheric pressure two times smaller, but this difference is likely insufficient to explain the discrepancy.  The day side of $\upsilon$~And~b receives 50\% more flux from its star, but it is unclear how this would affect the day/night temperature contrast.  Secondary eclipse depths for several planets have been in good agreement with the predictions from simple 1D models\cite{19,21,22,23} that assume a uniform day-night temperature, consistent with our conclusions for HD~189733b.  Taken together, these results argue for atmospheres that are very dark at visible wavelengths, likely absorbing 90\% or more of the incident stellar flux, and at the same time able to transport much of this energy to the night side.

\clearpage

\begin{addendum}

\item  We thank J. Winn for sharing data from a recent paper describing the behavior of the spots on the star, and D. Sasselov and E. Miller-Ricci for discussions on the properties of these spots.  This work is based on observations made with the Spitzer Space Telescope, which is operated by the Jet Propulsion Laboratory, California Institute of Technology, under contract to NASA.  Support for this work was provided by NASA through an award issued by JPL/Caltech.  We are grateful to the entire Spitzer team for their assistance throughout this process. H. K. was supported by a National Science Foundation Graduate Research Fellowship.

\item[Competing Interests] The authors declare that they have no competing 
financial interests.

\item[Correspondence] Requests for materials should be addressed to 
H.K.~(email: hknutson@cfa.harvard.edu).

\end{addendum}

\clearpage

\begin{figure}
\caption{Observed phase variation for HD 189733b, with transit and secondary eclipse visible.  We determine the location of the star by taking the weighted average of the flux contained in a 7-pixel box centered on the peak of the point spread function.  We find that the shape of the observed variation is consistent for apertures between $3.5-7$ pixels.  IRAC takes images in sets of 64, and we found that the average fluxes in the first three images and the 58th image were consistently low.  Additionally, 2\% of the images had corrupted pixels within our aperture.  We chose to trim both sets of images from our final time series.  We also exclude the first 1.8 hours of data from our analysis, as our correction was designed to correct the data only beyond the start of the transit.  We estimate the background flux by fitting a Gaussian function to a histogram of the fluxes from a subset of pixels located in the corners of the image.  This background contributes 1.3\% of the total flux in our aperture, and we subtract a constant value from our time series.  The scatter in the final timeseries is 20\% higher than predicted from photon noise alone; we use the standard deviation of the points after the end of the secondary eclipse as our error for each point.  The stellar flux as measured at the center of the secondary eclipse is normalized to unity (dashed line), and the data is binned every 500 points (200 s).  {\bf a} and {\bf b} show the same data, but in {\bf b} the y axis is expanded to show the scale of the variation.  The solid line is the phase curve for the best-fit model (Fig. 3).}
\end{figure}

\begin{figure}
\caption{Time series of the transit and secondary eclipse.  Data is binned every 100 points (40 s), with the out-of-transit fluxes normalized to unity. {\bf a} shows the transit and {\bf b} shows the secondary eclipse with best-fit eclipse curves overplotted, including timing offset, and residuals for the transit ({\bf c}) and secondary eclipse ({\bf d}) are plotted below.  The out of transit data for the eclipses are normalized using a constant; we find the transit occurs 1~s earlier and the secondary eclipse occurs 8~s earlier if we use a linear fit instead, an insignificant difference.  We fit both eclipses, fixing the mass of the star\cite{4} and allowing the transit times to vary freely\cite{11}.  From the primary eclipse, we find the radius of the planet is $1.137\pm0.006$~$R_{Jup}$, the orbital inclination is $85.61\pm0.04$~degrees, and the radius of the star is $0.757\pm0.003$~$R_{Sun}$; the planet-star radius ratio is $0.1545\pm0.0002$.  The formal uncertainty in the mass of the star introduces an additional error of $\pm1.8\%$ in our estimates for the two radii.  The depth of the secondary eclipse is $0.3381\pm0.0055$\% in relative flux.  Using a model\cite{12} we predict a stellar brightness temperature of 4512~K in the 8~$\mu$m Spitzer bandpass, where brightness temperature is defined by equating the Planck function with the mean surface brightness.  We note that our best-fit value for the depth of the transit (as characterized by ratio of the plantary to stellar radii) is slightly smaller than previous published values\cite{6}; this difference is most likely due to the effect of spots on the star.  Large spots would increase the apparent depth of the transit at visible wavelengths, while having a minimal impact at 8~$\mu$m.}
\end{figure}

\begin{figure}
\caption{Brightness estimates for 12 longitudinal strips on the surface of the planet.  Data are shown as a colour map ({\bf a}) and in graphical form ({\bf b}); see below for details.  We assume the planet is tidally locked, and we approximate it as being edge-on with no limb-darkening, so that the brightness of the $i$th slice is $F_i(\sin{\phi_{i,2}}-\sin{\phi_{i,1}})$ where $-\pi/2\le \phi_{i,1}, \phi_{i,2} \le \pi/2$ are the edges of the visible portion of each slice and $F_i$ is the flux from a slice when it is closest to us.  We bin the light curve into 32 bins with 4000 data points each, excising the data during the eclipses.  We define our goodness-of-fit parameter as $\chi^2 + \lambda \sum_{i=1}^{12} (F_i-F_{i-1})^2$, where $\chi^2$ is the goodness of fit for the light curve and the second term is a linear regularizing term which enforces small variations in adjacent slices for large $\lambda$ and allows a unique solution for $F_i$ for a given value of $\lambda$.  We optimize this function using a 1000-step Markov Chain Monte Carlo method to determine the planetary flux profile and corresponding uncertainties.  We chose a value for $\lambda$ that produced a reasonable compromise between the quality of the fit and the smoothness of the final brightness map.  We varied both the size of the bins and the number of longitudinal slices and our resulting slice fluxes are robust.  The brightness values in {\bf b} are given as the ratio of the flux from an individual slice viewed face-on to the total flux of the star, with 1$\sigma$ errors.  {\bf a} is a Mollweide projection of this brightness distribution, with an additional sinusoidal dependence on latitude included (note that the data provide no latitude information).  This plot uses a linear scale, with the brightest points in white and the darkest points in black.}
\end{figure}

\clearpage
\includegraphics[110,160][700,750]{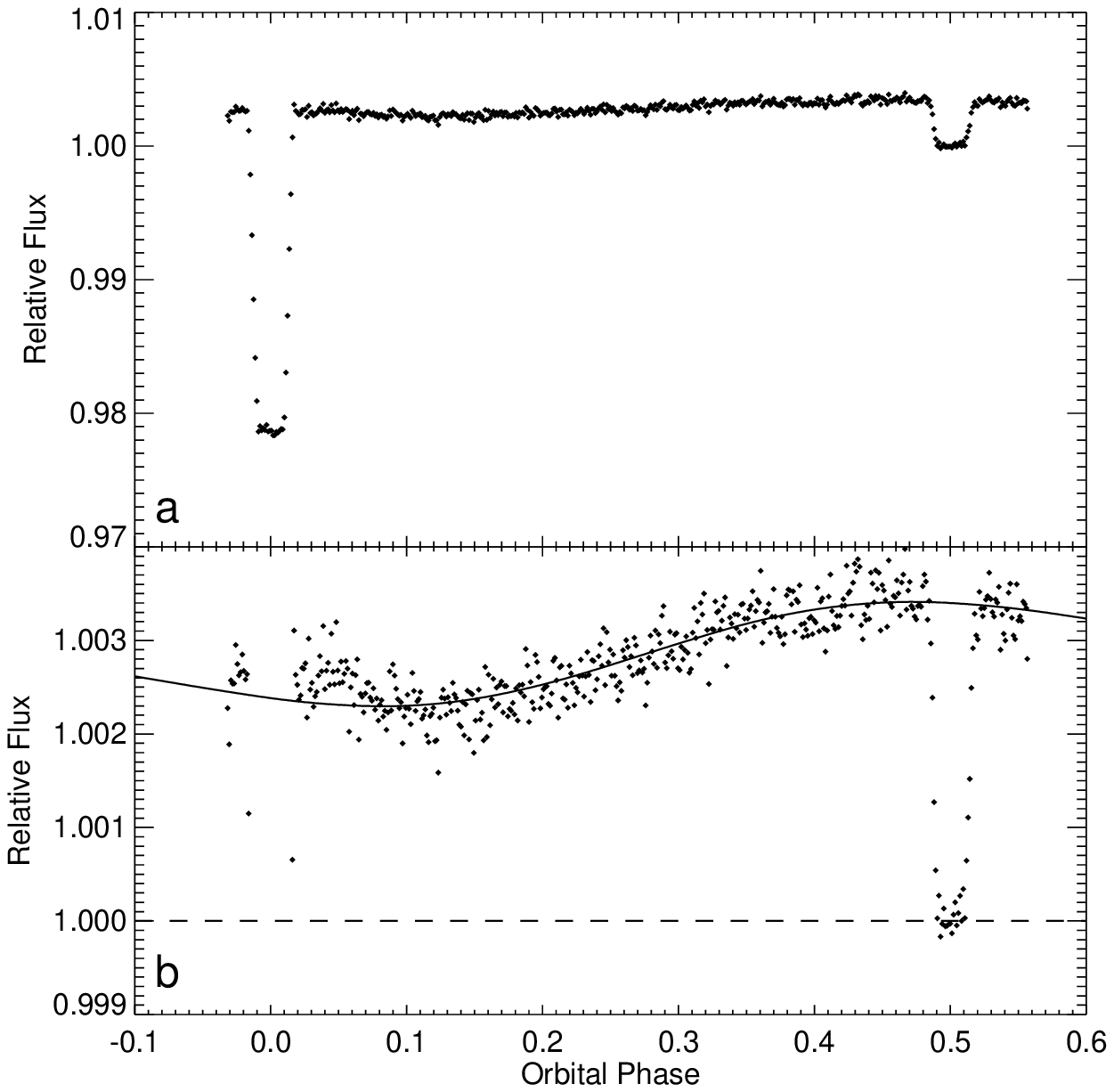}
\clearpage
\includegraphics[100,300][480,750]{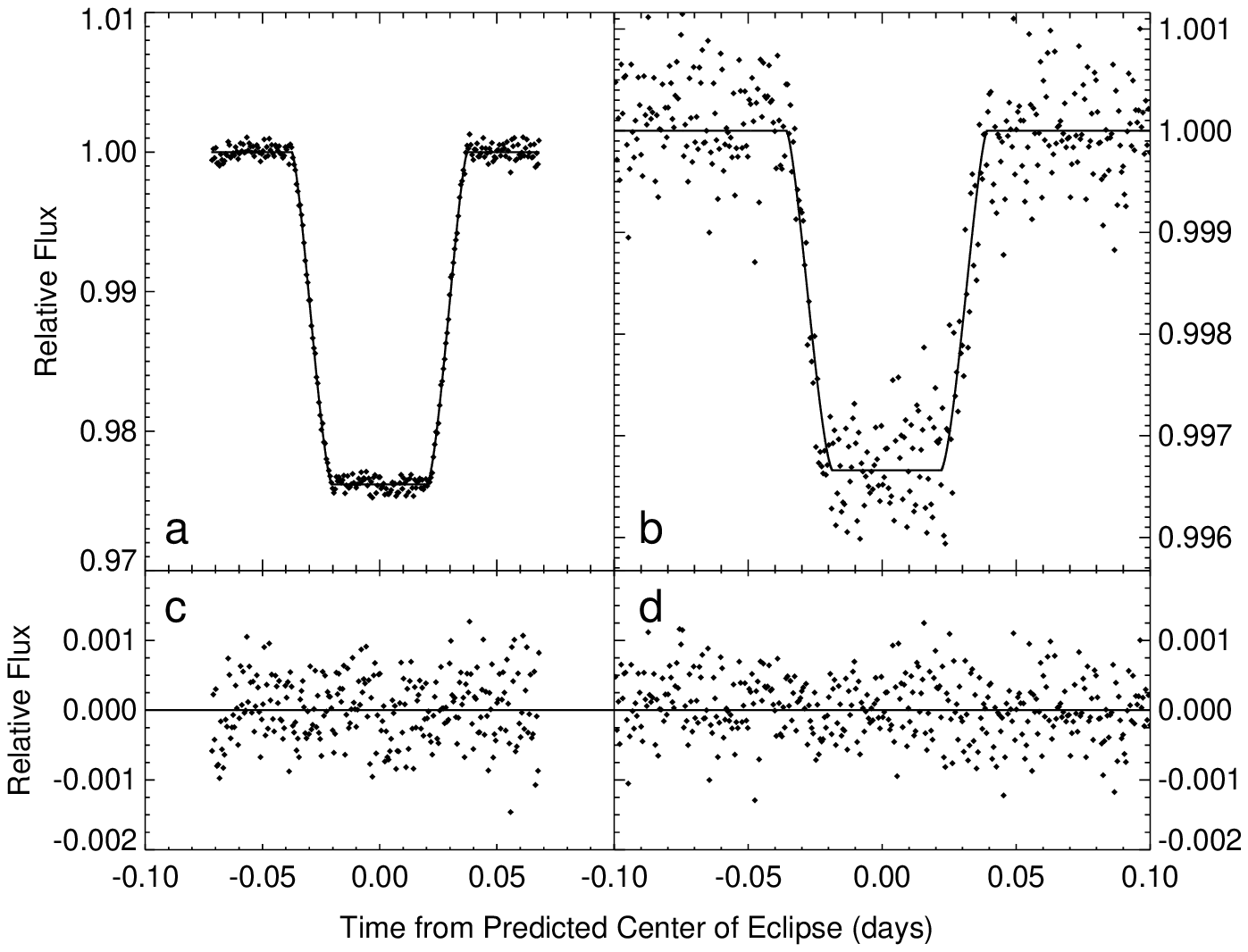}
\clearpage
\includegraphics{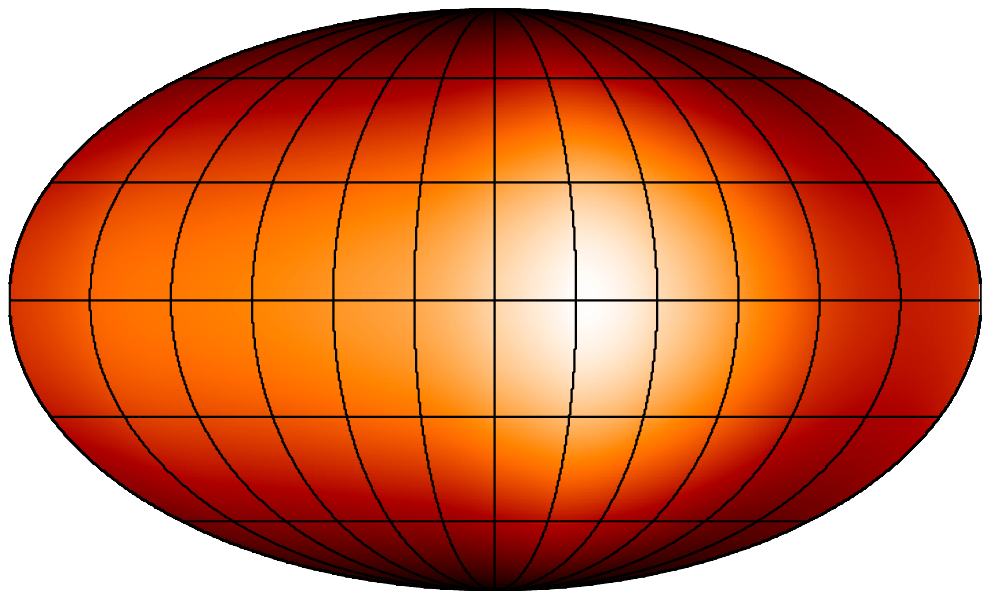}

\includegraphics[103,300][480,500]{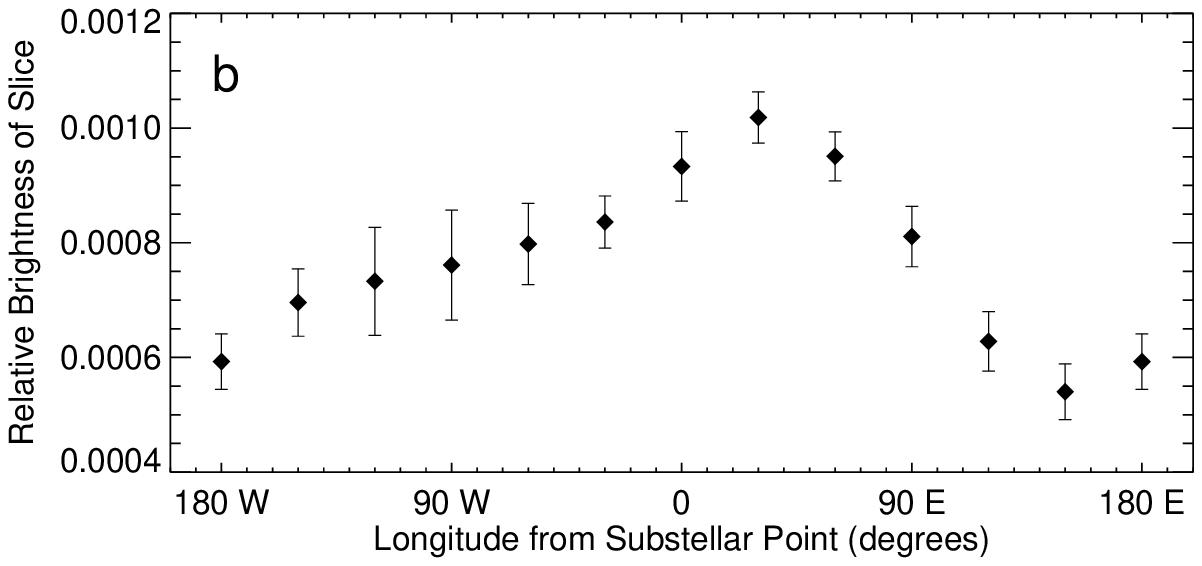}


\begin{thebibliography}{1}
\bibitem{1} Deming,~D., Seager,~S., Richardson,~L.~J., \& Harrington,~J. Infrared radiation from an extrasolar planet. \emph{Nature} {\bf 434}, 740-743 (2005).
\bibitem{2} Charbonneau,~D. et~al. Detection of thermal emission from an extrasolar planet. \emph{Astrophys.~J.} {\bf 626}, 523-529 (2005).
\bibitem{3} Deming,~D., Harrington,~J., Seager,~S., \& Richardson,~L.~J. Strong infrared emission from the extrasolar planet HD 189733b. \emph{Astrophys.~J.} {\bf 644}, 560-564 (2006).
\bibitem{4} Bouchy,~F.~et~al. ELODIE metallicity-biased search for transiting Hot Jupiters. II. A very hot Jupiter transiting the bright K star HD 189733. \emph{Astron. Astrophys.} {\bf 444}, L15-L19 (2005).
\bibitem{5} Bakos,~G.~A.~et~al. Refined parameters of the planet orbiting HD 189733. \emph{Astrophys.~J.} {\bf 650}, 1160-1171 (2006).
\bibitem{6} Winn,~J.~N.~et~al. The Transit Light Curve Project. V. System parameters and stellar rotation period of HD 189733. \emph{Astronom.~J.} {\bf 133}, 1828-1835 (2007).
\bibitem{7} Harrington,~J.~et~al. The phase-dependent infrared brightness of the extrasolar planet $\upsilon$ Andromeda b. \emph{Science} {\bf 314}, 5799, 623-626 (2006).
\bibitem{8} Fazio,~G.~G.~et~al. The Infrared Array Camera (IRAC) for the Spitzer Space Telescope.  \emph{Astrophys.~J.~Suppl.} {\bf 154}, 10-17 (2004).
\bibitem{9} Werner,~M.~W.~et~al. The Spitzer Space Telescope mission. \emph{Astrophys.~J.~Suppl.} {\bf 154}, 1-9 (2004).
\bibitem{10} Bakos,~G.~A., Andr\'as,~P., Latham,~D.~W., Noyes,~R.~W., Stefanik,~R.~P. A stellar companion in the HD 189733 system with a known transiting extrasolar planet. \emph{Astrophys.~J.} {\bf 641}, L57-L60 (2006).
\bibitem{11} Mandel,~K., \& Agol,~E. Analytic light curves for planetary transit searches. \emph{Astrophys.~J.} {\bf 580}, L171-L175 (2002).
\bibitem{12} Kurucz, R. \emph{Solar abundance model atmospheres for 0, 1, 2, 4, and 8 km/s.} CD-ROM 19 (Smithsonian Astrophysical Observatory, Cambridge, Massachusetts, 1994). 
\bibitem{13} Loeb,~A. A dynamical method for measuring the masses of stars with transiting planets. \emph{Astrophys.~J}. {\bf 623}, L45-L48 (2005).
\bibitem{14} Williams,~P.~K.~G., Charbonneau,~D., Cooper,~C.~S., Showman,~A.~P., \& Fortney,~J.~J. Resolving the surfaces of extrasolar planets with secondary eclipse light curves. \emph{Astrophys.~J.} {\bf 649}, 1020-1027 (2006).
\bibitem{15} Rauscher,~E.~et~al. Hot Jupiter variability in eclipse depth. \emph{ArXiv Astrophysics e-prints} (2007). \texttt{arXiv:astro-ph/0612412}
\bibitem{16} Bodenheimer,~P., Laughlin,~G., \& Lin,~D. On the radii of extrasolar giant planets. \emph{Astrophys.~J.} {\bf 592}, 555-563 (2003).
\bibitem{17} Guillot,~T., Burrow,~A., Hubbard,~W.~B., Lunine,~J.~I., \& Saumon,~D. Giant planets at small orbital distances. \emph{Astrophys.~J.} {\bf 459} L35-L38
\bibitem{18} Showman,~A.~P. \& Guillot,~T. Atmospheric circulation and tides of ``51 Pegasus b-like'' planets. \emph{Astron. Astrophys.} {\bf 385}, 166-180 (2002).
\bibitem{19} Seager,~S.~et~al. On the dayside thermal emission of hot Jupiters. \emph{Astrophys.~J.} {\bf 632}, 1122-1131 (2005).
\bibitem{20} Iro,~N., B\' ezard,~B., \& Guillot,~T. A time-dependent radiative model of HD 209458b. \emph{Astron. \& Astrophys.} {\bf 436}, 719-727 (2005).
\bibitem{21} Fortney,~J.~J., Marley,~M.~S., Lodders,~K., Saumon,~D., \& Freedman,~R. Comparative planetary atmospheres: models of TrES-1 and HD 209458b. \emph{Astrophys.~J.} {\bf 627}, L69-L72 (2005).
\bibitem{22} Barman,~T.~S., Hauschildt,~P.~H., \& Allard,~F. Phase-dependent properties of extrasolar planet atmospheres. \emph{Astrophys.~J.} {\bf 632}, 1132-1139 (2005).
\bibitem{23} Burrows,~A., Sudarsky,~D., \& Hubeny,~I. Theory for the secondary eclipse fluxes, spectra, atmospheres, and light curves of transiting extrasolar giant planets. Astrophys.~J. 650 1140-1149 (2006).
\bibitem{24} Cho,~J.~Y.-K., Menou,~K., Hansen,~B.~M.~S., \& Seager,~S. The changing face of the extrasolar giant planet HD 209458b. \emph{Astrophys.~J.} {\bf 587}, L117-L120 (2003).
\bibitem{25} Burkert,~A., Lin,~D.~N.~C., Bodenheimer,~P.~H., Jones,~C.~A., \& Yorke,~H.~W. On the surface heating of synchronously spinning short-period Jovian planets. \emph{Astrophys.~J.} {\bf 618}, 512-523 (2005).
\bibitem{26} Cooper,~C.~S., \& Showman,~A.~P. Dynamic meteorology at the photosphere of HD 209458b. \emph{Astrophys.~J.} {\bf 629}, L45-L48 (2005).
\bibitem{27} Cooper,~C.~S. \& Showman,~A.~P. Dynamics and disequilibrium carbon chemistry in hot Jupiter atmospheres, with application to HD 209458b. \emph{Astrophys.~J.} {\bf 649}, 1048-1063 (2006).
\bibitem{28} Langton,~J. \& Laughlin,~G. Observational consequences of hydrodynamic flows on hot Jupiters. \emph{Astrophys.~J.} {\bf 657} L113-L116 (2007).
\bibitem{29} Fortney,~J.~J., Saumon,~D., Marley,~M.~S., Lodders,~K., \& Freedman,~R.~S. Atmosphere, interior, and evolution of the metal-rich transiting planet HD 149026b. \emph{Astrophys.~J.} {\bf 642}, 495-504 (2006).
\bibitem{30} Fortney,~J.~J., Cooper,~C.~S., Showman,~A.~P., Marley,~M.~S., \& Freedman,~R.~S. The influence of atmospheric dynamics on the infrared spectra and light curves of hot Jupiters. \emph{Astrophys.~J.} {\bf 652}, 746-757 (2006).
\end{thebibliography}
\end{document}